\documentclass[12pt]{article}
\usepackage{latexsym}
\usepackage{exscale}
\usepackage{psfig}
\begin{document}

\begin{center}
\LARGE
Interaction of the quantized electromagnetic field with atoms 
in the presence of dispersing and absorbing dielectric bodies\\[2ex]
\large
S.~Scheel and D.-G.~Welsch\\[1ex]
Theoretisch-Physikalisches Institut,\\
Friedrich-Schiller-Universit\"{a}t Jena,\\
Max-Wien-Platz 1, D-07743 Jena, Germany\\[4ex]
\end{center}
\thispagestyle{empty}

\begin{abstract}
A general theory of the interaction of the quantized 
electromagnetic field with atoms in the presence of dispersing 
and absorbing dielectric bodies of given Kramers--Kronig consistent 
permittivities is developed. It is based on a source-quantity 
representation of the electromagnetic field, in which the 
electromagnetic-field operators are expressed in terms of a continuous
set of fundamental bosonic fields via the Green tensor of the classical
problem. Introducing scalar and vector potentials, the formalism is 
extended in order to include in the theory the interaction of the 
quantized electromagnetic field with additional atoms. Both the 
minimal-coupling scheme and the multipolar-coupling scheme are 
considered. The theory replaces the standard concept of mode 
decomposition which fails for complex permittivities. It enables 
us to treat the effects of dispersion and absorption 
in a consistent way and to give a unified approach to the
atom-field interaction, without any restriction to a particular
interaction regime in a particular frequency range. All relevant 
information about the dielectric bodies such as form and intrinsic 
dispersion and absorption is contained in the Green tensor.
The application of the theory to the spontaneous 
decay of an excited atom in the presence of dispersing 
and absorbing bodies is addressed.  
\end{abstract}


\section{Introduction}

Optical instruments such as beam splitters or cavities are 
more or less complicated macroscopic material bodies, whose
use in experiments requires careful examination with regard to their action
on the light under study. In quantum optics an important 
consideration is the influence of the presence of 
material bodies on the quantum statistics of the light. It is 
therefore necessary to take account of the presence of material
bodies when considering the interaction of quantized light with
atomic systems. In principle, such bodies could be included as a part 
of the matter to which the radiation field is coupled and treated 
microscopically. However there is a class of material bodies whose
action can be included in the quantum theory exactly,
namely dielectric bodies that respond linearly to the electromagnetic
field.

The quantum theory of radiation in the presence of dielectric media
has been studied over a long period. Commonly, dielectric matter is 
characterized by the permittivity, which describes
the response of the matter to the electric field.
Quantization of the electromagnetic field in dielectrics with
real and frequency-independent permittivity has been treated extensively
\cite{Jauch48,Shen67,Carniglia71,Birula72,Abram87,Birula87,%
Knoll87,Kennedy88,Khosravi91,Glauber91,Dalton96,Bordag98,Dalton99}. 
In the same context, dispersive dielectrics have been considered 
\cite{Watson49,Agarwal75,Huttner91,Drummond91,Milonni95,Santos95,Drummond99}.
However, it is well known that the permittivity is a complex function of 
frequency which has to satisfy the Kramers--Kronig relations which state 
that the real part of the permittivity (responsible for dispersion) and 
the imaginary part (responsible for absorption) 
are necessarily connected with each other. 
Hence, any quantum theory that is based
on the assumption of a real permittivity can only be valid
for narrow-bandwidth fields far from medium resonances 
where absorption can safely be disregarded.

A systematic and quantum-theoretically consistent approach 
to the problem was developed by Huttner and Barnett 
\cite{Huttner92} on the basis of the microscopic Hopfield model 
of a bulk dielectric \cite{Hopfield58}. 
They performed an explicit Fano-type diagonalization \cite{Fano56} of a
Hamiltonian consisting of the electromagnetic field, a (harmonic-oscillator)
polarization field representing the dielectric matter, and a continuous set 
of (harmonic-oscillator) reservoir variables accounting for absorption. 
The resulting expression for the vector potential could also be 
written in terms of the Green function of the classical scattering 
problem as was realized by Gruner and Welsch
\cite{Gruner96a,Gruner96c}. That, in fact, makes it possible to perform
the quantization of the electromagnetic field in the presence
of arbitrary dielectric bodies of phenomenologically given 
Kramers--Kronig consistent permittivities, without referring to 
specific microscopic models of the bodies, which are hard to 
establish for general systems \cite{Ho98,Scheel98}. 

Having quantized the electromagnetic field, 
the question arises of how to include in the
theory the interaction of the medium-assisted field with
atomic systems in order to study the influence of 
dielectric bodies on quantum-optical processes such
as spontaneous decay of an excited atom near material
bodies or in micro-cavities. It is well known that
the spontaneous decay can be strongly modified when
the atom is placed inside a high-quality micro-cavity. 
Recent progress in constructing micro-cavities has rendered
it possible to approach the ultimate quality level 
determined by intrinsic material losses, so that the 
question of the influence of absorbing material has been 
of increasing interest. In what follows we develop
a basic-theoretical concept for treating the
interaction of the electromagnetic field with
atomic systems in the presence of dielectric bodies
of (phenomenologically) given arbitrary complex permittivities.  

The paper is organized as follows. In Sec.~\ref{quantization} 
the quantization of the electromagnetic field on a dielectric
background of complex permittivity is outlined.
The interaction of the medium-assisted electromagnetic field 
with atomic systems is studied in Sec.~\ref{interaction}
and the minimal-coupling and multipolar-coupling
Hamiltonians are derived. Finally, a summary
and some concluding remarks are given in Sec.~\ref{summary}.


\section[Quantization of the electromagnetic field ]{Quantization of the 
electromagnetic field in the presence of dielectric media}
\label{quantization}

As already mentioned, dielectric matter plays an
important role in optics, because (passive) optical
instruments are typically composed of dielectrics.
In classical electrodynamics, dielectric matter 
is commonly described in terms of a phenomenologically introduced
dielectric susceptibility (or permittivity).
This concept has the benefit of being universally
valid, because it uses only general physical 
properties, without the need of involved 
{\em ab initio} calculations. 

\subsection{Classical basic equations}

The phenomenological Maxwell equations of the electromagnetic field 
in the presence of dielectric bodies but without additional charge 
and current densities read   
\begin{equation}
\label{2.01}
\mbox{\boldmath $\nabla$}\cdot{\bf B}({\bf r}) = 0,
\end{equation}
\begin{equation}
\label{2.02}
\mbox{\boldmath $\nabla$}\times{\bf E}({\bf r})
+ \dot {\bf B}({\bf r}) =0,             
\end{equation}
\begin{equation}
\label{2.03}
\mbox{\boldmath $\nabla$}\cdot{\bf D}({\bf r}) = 0,
\end{equation}
\begin{equation}
\label{2.04}
\mbox{\boldmath $\nabla$}\times{\bf H}({\bf r})
- \dot {\bf D}({\bf r}) =0,                   
\end{equation}
where the displacement field ${\bf D}$ is related to the
electric field ${\bf E}$ and the polarization field ${\bf P}$ 
according to
\begin{equation}
\label{2.05}
{\bf D}({\bf r}) = \varepsilon_0 {\bf E}({\bf r}) 
+ {\bf P}({\bf r}), 
\end{equation}
and for nonmagnetic matter it may be assumed that
\begin{equation}
\label{2.06}
{\bf H}({\bf r}) = \frac{1}{\mu_0}\, {\bf B}({\bf r}).
\end{equation}

Let us restrict our attention
to isotropic but arbitrarily inhomogeneous media 
and assume that the polarization linearly and locally responds
to the electric field. In this case, the most general
relation between the polarization and the electric field
which is in agreement with the causality principle and the
dissipation-fluctuation theorem is
\begin{equation}
\label{2.07}
{\bf P}({\bf r},t) 
= \varepsilon_0 \int\limits_0^\infty {\rm d}\tau \, \chi({\bf r},\tau)
{\bf E}({\bf r},t-\tau) + {\bf P}_{\rm n}({\bf r},t),
\label{2.4.6}
\end{equation} 
where $\chi({\bf r},\tau)$ is the dielectric susceptibility
as a function of space and time, and ${\bf P}_{\rm n}$ is the
(noise) polarization associated with absorption. 

Substitution of this expression into 
Eq.~(\ref{2.05}) together with Fourier transformation converts 
this equation to
\begin{equation}
\label{2.08}
\underline{\bf D}({\bf r},\omega) 
= \varepsilon_0 \varepsilon({\bf r},\omega)
\underline{\bf E}({\bf r},\omega) 
+ \underline{\bf P}_{\rm n}({\bf r},\omega), 
\label{2.4.7}
\end{equation}
where
\begin{equation}
\label{2.09}
\varepsilon({\bf r},\omega)
= 1 + \int\limits_0^\infty {\rm d}\tau \, \chi({\bf r},\tau) e^{i\omega\tau}
\end{equation}
is the (relative) permittivity,
and the Maxwell equations (\ref{2.01})--(\ref{2.04}) read in 
the Fourier domain as
\begin{equation}
\label{2.13}
\mbox{\boldmath $\nabla$}\cdot\underline{\bf B}({\bf r},\omega)=0, 
\end{equation}
\begin{equation}
\label{2.14}
\mbox{\boldmath $\nabla$}\times\underline{\bf E}({\bf r},\omega)
= i\omega \underline{\bf B}({\bf r},\omega),
\end{equation}
\begin{equation}
\label{2.15}
\varepsilon_0 \mbox{\boldmath $\nabla$}\cdot
\varepsilon({\bf r},\omega) \underline{\bf E}({\bf r},\omega)
= \underline{\rho}_{\rm n}({\bf r},\omega),  
     \end{equation}
\begin{equation}
\label{2.16}
\mbox{\boldmath $\nabla$}\times\underline{\bf B}({\bf r},\omega)
+i\frac{\omega}{c^2} \varepsilon({\bf r},\omega) 
\underline{\bf E}({\bf r},\omega)
= \mu_0 \underline{\bf j}_{\rm n}({\bf r},\omega) 
\end{equation}
($c^{-2}=\varepsilon_0 \mu_0$ ).\footnote{Here and in 
   the following the Fourier transform
   $\underline{F}(\omega)$ of a real function $F(t)$ is defined
   according to the relation 
   $F(t)$ $\!=$ $\!\int_0^\infty {\rm d}\omega\, 
   \underline{F}(\omega) e^{-i\omega t}$ $\!+$ $\!{\rm c.c.}$.}
Here we have introduced the noise charge density
\begin{equation}
\label{2.17}
\underline{\rho}_{\rm n}({\bf r},\omega)
= - \mbox{\boldmath $\nabla$}\cdot
\underline{\bf P}_{\rm n}({\bf r},\omega)
\end{equation}
and the noise current density
\begin{equation}
\label{2.18}
\underline{\bf j}_{\rm n}({\bf r},\omega)
= - i\omega \underline{\bf P}_{\rm n}({\bf r},\omega),
\end{equation}
which obey the continuity equation 
\begin{equation}
\label{2.19}
\mbox{\boldmath $\nabla$}\cdot
\underline{\bf j}_{\rm n}({\bf r},\omega) \!= 
\!i\omega \underline{\rho}_{\rm n}({\bf r},\omega).
\end{equation}

According to Eq.~(\ref{2.09}), the permittivity $\epsilon({\bf r},\omega)$
is a complex function of frequency,
\begin{equation}
\label{2.19a}
\epsilon({\bf r},\omega) 
= \epsilon'({\bf r},\omega)
 + i\,\epsilon''({\bf r},\omega).
\end{equation}
The real and imaginary parts, which are responsible
for dispersion and absorption respectively, are uniquely
related to each other through the Kramers--Kronig relations
\begin{equation}
\label{2.11}
\varepsilon'({\bf r},\omega)
- 1 = \frac{{\cal P}}{\pi} 
\int {\rm d}\omega' \, \frac{
\varepsilon''({\bf r},\omega')
}{\omega'-\omega}\,, 
\label{2.4.9}
\end{equation}
\begin{equation}
\label{2.12}
\varepsilon''({\bf r},\omega)  
= - \frac{{\cal P}}{\pi}  \int {\rm d}\omega' \,
\frac{\varepsilon'({\bf r},\omega') -1}{\omega'-\omega} 
\end{equation}
(${\cal P}$, principal value). Further,
$\varepsilon({\bf r},\omega)$ as a function of complex $\omega$
satisfies the relation 
\begin{equation}
\label{2.12a}
\varepsilon({\bf r},-\omega^\ast)
= \varepsilon^\ast({\bf r},\omega) 
\end{equation}
and is holomorphic in the upper complex half-plane without 
zeros. In particular, it approaches 
unity in the high-frequency limit, i.e, 
$\varepsilon({\bf r},\omega)$ $\!\to$ $\!1$ if
\mbox{$|\omega|$ $\!\to$ $\!\infty$}.

The Maxwell equations (\ref{2.14}) and 
(\ref{2.16}) imply that  
$\underline{\bf E}({\bf r},\omega)$ 
obeys the partial differential equation 
\begin{equation}
\label{2.20}
\mbox{\boldmath $\nabla$}\times  
\mbox{\boldmath $\nabla$}\times
\underline{\bf E}({\bf r},\omega) 
-\frac{\omega^2}{c^2} \varepsilon({\bf r},\omega) 
\underline{\bf E}({\bf r},\omega) \!
= \! i\omega \mu_0 \underline{\bf j}_{\rm n}({\bf r},\omega),
\end{equation}
whose solution can be represented in the form
\begin{equation}
\label{2.21}
\underline{\bf E}({\bf r},\omega)
= i\mu_0\omega \int {\rm d}^3{\bf r}' \,
\mbox{\boldmath $G$}({\bf r},{\bf r}',\omega)
\cdot \underline{\bf j}_{\rm n}({\bf r}',\omega),
\end{equation}
where the Green tensor  
$\mbox{\boldmath $G$}({\bf r},{\bf r}',\omega)$
has to be determined from the equation
\begin{equation}
\label{2.22}
\mbox{\boldmath $\nabla$}\times  
\mbox{\boldmath $\nabla$}\times
\mbox{\boldmath $G$}({\bf r},{\bf r}',\omega) 
-\frac{\omega^2}{c^2} \varepsilon({\bf r},\omega) 
\mbox{\boldmath $G$}({\bf r},{\bf r}',\omega)
= \mbox{\boldmath $\delta$}({\bf r}-{\bf r}')
\end{equation}
together with the boundary condition at infinity.
In Cartesian coordinates, Eq.(\ref{2.22}) reads
\begin{equation}
\label{2.22a}
\left[ \left(\partial^r_i \partial^r_m - \delta_{im} 
\Delta^r\right)- \delta_{im} {\omega^2\over c^2}\varepsilon({\bf
r},\omega) \right] 
G_{mj}({\bf r}, {\bf r}',\omega) 
= \delta_{ij} \delta({\bf r}-{\bf r}')  
\end{equation}
($\partial^r_i$ $\!=$ $\!\partial/\partial x_i$), where over repeated
vector-component indices is summed.  The Green tensor has the
properties that
\begin{equation}
\label{2.22b}
G_{ij}({\bf r},{\bf r}',\omega)^\ast =  G_{ij}({\bf r},{\bf r}',-\omega),
\end{equation}
\begin{equation}
\label{2.22c}
G_{ji}({\bf r}',{\bf r},\omega) =  G_{ij}({\bf r},{\bf r}',\omega),
\end{equation}
and
\begin{equation}
\label{2.22d}
   \int {\rm d}^3{\bf s} \, \frac{\omega^2}{c^2} 
   \varepsilon''({\bf s},\omega)\, G_{ik}({\bf r},{\bf s},\omega)     
   G^*_{jk}({\bf r'},{\bf s},\omega) = {\rm Im} \, 
   G_{ij}({\bf r},{\bf r'},\omega). 
\end{equation} 
The property (\ref{2.22b}) is a direct consequence of the corresponding
relation (\ref{2.12a}) for the permittivity, Eq.~(\ref{2.22c}) is the 
so-called reciprocity relation, and (\ref{2.22d}) is proved in
\cite{Scheel98}. 

The Fourier components of the magnetic induction, 
$\underline{\bf B}({\bf r},\omega)$, and the  displacement field,
$\underline{\bf D}({\bf r},\omega)$, are directly related to the
Fourier components of the
electric field, $\underline{\bf E}({\bf r},\omega)$,   
\begin{equation}
\label{2.23}
\underline{\bf B}({\bf r},\omega)
= (i\omega)^{-1}\mbox{\boldmath $\nabla$}\times
\underline{\bf E}({\bf r},\omega), 
\end{equation}
\begin{equation}
\label{2.24}
\underline{\bf D}({\bf r},\omega)
= (\mu_0\omega^2)^{-1} \mbox{\boldmath $\nabla$}\times
\mbox{\boldmath $\nabla$} \times \underline{\bf E}({\bf r},\omega)
\end{equation}
[see Eqs.~(\ref{2.14}), (\ref{2.08}), (\ref{2.18}), and (\ref{2.20})], 
and $\underline{\bf E}({\bf r},\omega)$ is determined, according to
Eq.~(\ref{2.21}), by ${\bf j}_{\rm n}({\bf r},\omega)$.
The continuous set of (complex) fields  
${\bf j}_{\rm n}({\bf r},\omega)$ [or, equivalently 
${\bf P}_{\rm n}({\bf r},\omega)$] can therefore be regarded as playing 
the role of the set of dynamical variables of the system composed of the
electromagnetic field and the medium (including the dissipative system). 
For the following it is convenient to split off some factor
from ${\bf P}_{\rm n}({\bf r},\omega)$ and to define
the fundamental dynamical variables ${\bf f}({\bf r},\omega)$ 
as follows:
\begin{equation}
\label{2.25}
\underline{\bf P}_{\rm n}({\bf r},\omega)
= i\sqrt{\frac{\hbar\varepsilon_0}{\pi}\,
\varepsilon''({\bf r},\omega) 
} \,  {\bf f}({\bf r},\omega).
\end{equation}


\subsection{Field quantization}

The transition from classical to quantum theory now consists
in the replacement of the classical fields 
${\bf f}({\bf r},\omega)$ and ${\bf f}^\ast({\bf r},\omega)$
by the operator-valued bosonic fields 
$\hat{\bf f}({\bf r},\omega)$ and $\hat{\bf f}^\dagger({\bf r},\omega)$
respectively, which are associated with the
elementary excitations of the 
composed system in linear approximation. Thus the commutation relations are
\begin{equation}
\label{2.26}
\big[\hat{f}_k({\bf r},\omega),\hat{f}_{k'}^\dagger({\bf r}',\omega')\big]
= \delta_{kk'} \delta({\bf r}\!-\!{\bf r}')\delta(\omega\!-\!\omega'), 
\end{equation}
\begin{equation}
\label{2.27}
\big[\hat{f}_k({\bf r},\omega),\hat{f}_{k'}({\bf r}',\omega')\big] = 0, 
\end{equation}
and the Hamiltonian of the composed system is
\begin{equation}
\label{2.28}
\hat{H} = \int {\rm d}^3{\bf r}\int\limits_0^\infty {\rm d}\omega
\,\hbar \omega\, \hat{\bf f}^\dagger({\bf r},\omega)\!\cdot\!
\hat{\bf f}({\bf r},\omega) 
\end{equation}

Recalling Eqs.~(\ref{2.18}) and (\ref{2.25}) and replacing  
${\bf E}({\bf r},\omega)$ [Eq.~(\ref{2.21})], 
${\bf B}({\bf r},\omega)$ [Eq.~(\ref{2.23})], and 
${\bf D}({\bf r},\omega)$ [Eq.~(\ref{2.24})]
by the quantum-mechanical operators, we find that
\begin{equation}
\label{2.29}
\underline{\hat{\bf E}}({\bf r},\omega)
= i \sqrt{\frac{\hbar}{\pi\varepsilon_0}}\,\frac{\omega^2}{c^2} 
\int {\rm d}^3{\bf r}' \sqrt{
\varepsilon''({\bf r}',\omega)
}\,
\mbox{\boldmath $G$}({\bf r},{\bf r}',\omega)
\cdot \hat{\bf f}({\bf r}',\omega),
\end{equation}
\begin{equation}
\label{2.30}
\underline{\hat{\bf B}}({\bf r},\omega)
= (i\omega)^{-1} \mbox{\boldmath $\nabla$}\times
\underline{\hat{\bf E}}({\bf r},\omega), 
\end{equation}
and 
\begin{equation}
\label{2.31}
\underline{\hat{\bf D}}({\bf r},\omega)
= (\mu_0\omega^2)^{-1} \mbox{\boldmath $\nabla$}\times
\mbox{\boldmath $\nabla$} \times 
\underline{\hat{\bf E}}({\bf r},\omega),
\end{equation}
from which the electromagnetic field operators in the Schr\"{o}dinger
picture are obtained by integration over $\omega$:
\begin{equation}
\label{2.32}
\hat{\bf E}({\bf r}) \equiv \hat{\bf E}_{\rm m}({\bf r}) 
= \int\limits_0^\infty {\rm d}\omega \, \underline{\hat{\bf E}}({\bf r},\omega)
+ {\rm H.c.},
\end{equation}
\begin{equation}
\label{2.33}
\hat{\bf B}({\bf r})
= \int\limits_0^\infty {\rm d}\omega \, \underline{\hat{\bf B}}({\bf r},\omega)
+ {\rm H.c.},
\end{equation}
and\footnote{The longitudinal (${\bf F}^\|$) and transverse
   (${\bf F}^\perp$) parts of a vector field ${\bf F}$ are defined by
   ${\bf F}^{\|(\perp)}({\bf r})$ $\!=$ $\!\int {\bf d}^3{\bf r}'$
   $\!\mbox{\boldmath $\delta$}^{\|(\perp)}({\bf r}-{\bf r}')
   {\bf F}({\bf r}')$,  
   with $\mbox{\boldmath $\delta$}^{(\|)}({\bf r})$ and 
   $\mbox{\boldmath $\delta$}^{(\perp)}({\bf r})$ being
   the longitudinal and transverse tensor-valued $\delta$-functions
   respectively.} 
\begin{equation}
\label{2.34}
\hat{\bf D}({\bf r}) \equiv \hat{\bf D}_{\rm m}({\bf r}) 
= \hat{\bf D}_{\rm m}^\perp({\bf r}) 
= \int\limits_0^\infty {\rm d}\omega \, \underline{\hat{\bf D}}({\bf r},\omega)
+ {\rm H.c.}.
\end{equation}
In this way, the electromagnetic field is expressed in terms of
the classical Green tensor $\mbox{\boldmath $G$}({\bf r},{\bf r}',\omega)$
satisfying the generalized Helmholtz equation (\ref{2.22})
and the continuum of the fundamental bosonic field variables 
$\hat{\bf f}({\bf r},\omega)$ [and $\hat{\bf f}^\dagger({\bf r},\omega)$]. 
All the information about the dielectric matter (such as its formation
in space and its dispersive and absorptive properties) 
is contained [via the permittivity $\varepsilon({\bf r},\omega)$] 
in the Green tensor of the classical problem.
Eqs.~(\ref{2.32}) -- (\ref{2.34}), together with
Eqs.~(\ref{2.29}) -- (\ref{2.31})   
can be considered as the generalization of the familiar mode decomposition.

The quantization scheme meets the basic requirements of quantum
electrodynamics. So it can be shown
by using very general properties of the permittivity and the Green tensor
that $\hat{\bf E}$ and $\hat{\bf B}$ satisfy the correct (equal-time)
commutation relations \cite{Scheel98}
\begin{equation}
\label{2.35}
\big[\hat{E}_k({\bf r}),\hat{E}_{k'}({\bf r}')\big] = 0 =
\big[\hat{B}_k({\bf r}),\hat{B}_{k'}({\bf r}')\big], 
\end{equation}
\begin{equation}
\label{2.36}
\big[\varepsilon_0\hat{E}_k({\bf r}),\hat{B}_{k'}({\bf r}')\big] = 
-i \hbar \,\epsilon_{kk'l} \,\partial^r_l \delta({\bf r}-{\bf r}'). 
\end{equation}
Obviously, the electromagnetic field operators in the Heisenberg picture 
satisfy the Maxwell equations (\ref{2.01}) -- (\ref{2.04}), 
with the time derivative of any operator $\hat{Q}$ being given by 
\begin{equation}
\label{2.37}
\dot{\hat{Q}} = (i \hbar)^{-1} \big[\hat{Q},\hat{H}\big],
\end{equation}
where $\hat{H}$ is the Hamiltonian (\ref{2.28}), 

Further, scalar ($\hat{\varphi}$) and vector ($\hat{\bf A}$) 
potentials can be introduced and expressed in terms
of the fundamental bosonic fields. 
In particular, the potentials in the Coulomb gauge are defined 
by
\begin{equation}
\label{2.39}
-\mbox{\boldmath $\nabla$}\hat{\varphi}({\bf r}) 
= \hat{\bf E}^\parallel({\bf r}) , 
\end{equation}
\begin{equation}
\label{2.38}
\hat{\bf A}({\bf r}) = 
\int\limits_0^\infty {\rm d} \omega\,
\underline{\hat{\bf A}}({\bf r},\omega)
+ {\rm H.c.} \ ,
\end{equation}
where
\begin{equation}
\label{2.38a}
\underline{\hat{\bf A}}({\bf r},\omega) 
= (i\omega)^{-1} \underline{\hat{\bf E}}^\perp({\bf r},\omega) 
\end{equation}
The canonically conjugated momentum field with respect to 
$\hat{\bf A}({\bf r})$ is  
\begin{equation}
\label{2.41}
\hat{\bf \Pi}({\bf r})
= -i\varepsilon_0 \int\limits_0^\infty {\rm d}\omega \, 
\omega \underline{\hat{\bf A}}({\bf r},\omega)
+ {\rm H.c.},
\end{equation}
and it is not difficult to verify that $\hat{\bf \Pi}$
$\!=$ $\!-\varepsilon_0\hat{\bf E}^\perp$,
$\mbox{\boldmath $\nabla$}\times\hat{\bf A}$ 
$\!=$ $\!\hat{\bf B}$, and
$-\dot{\hat{\bf A}}$ $\!-$ 
$\!\mbox{\boldmath $\nabla$}\hat{\varphi}({\bf r})$ $\!=$ $\!\hat{\bf E}$. 
In addition, $\hat{\bf A}$ and $\hat{\bf \Pi}$
satisfy the well-known commutation relations
\begin{equation}
\label{2.42}
\big[\hat{A}_k({\bf r}),\hat{A}_{k'}({\bf r}')\big] = 0 =
\big[\hat{\Pi}_k({\bf r}),\hat{\Pi}_{k'}({\bf r}')\big], 
\end{equation}
\begin{equation}
\label{2.43}
\big[\hat{A}_{k}({\bf r}),\hat{\Pi}_{k'}({\bf r}')\big] = 
i \hbar \,\delta^{\perp}_{kk'}({\bf r}-{\bf r}'). 
\end{equation}


\section[Interaction with charged particles]{Interaction of the
medium-assisted quantized electromagnetic field with charged particles}
\label{interaction}

The interaction of the quantized electromagnetic field with atoms placed
inside a dielectric medium or near dielectric bodies can be strongly
influenced by the dielectric medium. A well-known example is the dependence of
the spontaneous decay rate of an excited atom on the properties of
an dielectric environment. In order to study such and related
phenomena, the Hamiltonian (\ref{2.28}) must be supplemented with the 
the Hamiltonian of additional charged particles and
their interaction energy with the medium-assisted electromagnetic field. 
 
\subsection{The minimal-coupling Hamiltonian}

Applying the minimal-coupling scheme, we may write the
the total Hamiltonian in the form
\begin{eqnarray}
\label{2.51}
\lefteqn{
\hspace{-4ex}
\hat{H} = \int {\rm d}^3{\bf r} \int\limits_0^\infty {\rm d}\omega\,
 \hbar\omega\,\hat{\bf f}^\dagger({\bf r},\omega) \!\cdot\!
 \hat{\bf f}({\bf r},\omega) + \sum_\alpha {1\over 2m_\alpha}
  \left[ \hat{\bf p}_\alpha - q_\alpha 
  \hat{\bf A}(\hat{\bf r}_\alpha) \right]^2    
}
\nonumber\\&&\hspace{8ex}
+{\textstyle\frac{1}{2}} \int {\rm d}^3{\bf r}\, 
\hat{\rho}_{\rm A}({\bf r}) \hat{\varphi}_{\rm A}({\bf r})
+ \int {\rm d}^3{\bf r}\, 
\hat{\rho}_{\rm A}({\bf r}) \hat{\varphi}({\bf r}) ,
\end{eqnarray}
where $\hat{\bf r}_\alpha$ is the position operator and $\hat{\bf p}_\alpha$ 
is the canonical momentum operator of the $\alpha$th
(non-relativistic) particle of  charge $q_\alpha$ and mass $m_\alpha$.
The Hamiltonian (\ref{2.51}) consists of four terms. The first term is the
energy of the electromagnetic field and the medium (including the
dissipative system), as introduced in Eq.~(\ref{2.28}).
The second term is the kinetic energy of the charged particles, and the third
term is their Coulomb energy, where the corresponding scalar potential
$\hat{\varphi}_{\rm A}$ is given by 
\begin{equation}
\label{2.52}
\hat{\varphi}_{\rm A}({\bf r}) = 
\int {\rm d}^3{\bf r}' \frac {\hat{\rho}_{\rm A}({\bf r}')}
{4\pi\epsilon_0|{\bf r}-\hat{\bf r}'|} \,,
\end{equation}
with
\begin{equation}
\label{2.53}
\hat{\rho}_{\rm A}({\bf r}) = 
\sum_\alpha q_\alpha 
\delta({\bf r}-\hat{\bf r}_\alpha) 
\end{equation}
being the charge density. The last term is the Coulomb energy of interaction
of the particles with the medium.
{F}rom Eq.~(\ref{2.51}) it follows that 
the interaction Hamiltonian reads  
\begin{equation}
\label{2.63}
\hat{H}_{\rm int} = 
- \sum_\alpha {1\over m_\alpha} \left[ \hat{\bf p}_\alpha 
- {\textstyle{1\over 2}} q_\alpha \hat{\bf A}(\hat{\bf r}_\alpha)
\right]q_\alpha \hat{\bf A}(\hat{\bf r}_\alpha)
+ \int {\rm d}^3{\bf r}\, 
\hat{\rho}_{\rm A}({\bf r}) \hat{\varphi}({\bf r}).
\end{equation}
Note that in Eq.~(\ref{2.63})
the scalar potential $\hat{\varphi}$
and the vector potential $\hat{\bf A}$ must be thought of as being expressed,
on using Eqs.~(\ref{2.39}), (\ref{2.38}), (\ref{2.38a}) together
with Eqs.~(\ref{2.29}) and (\ref{2.32}),
in terms of the fundamental fields $\hat{\bf f}({\bf r},\omega)$
and $\hat{\bf f}^\dagger({\bf r},\omega)$.

In a straightforward but somewhat lengthy calculation 
it can be shown that both the operator-valued Maxwell equations 
\begin{equation}
\label{2.54}
\mbox{\boldmath $\nabla$}\cdot\hat{{\bf B}}({\bf r}) = 0,
\end{equation}
\begin{equation}
\label{2.55}
\mbox{\boldmath $\nabla$}\times\hat{{\bf E}}({\bf r})
+ \dot{\hat{{\bf B}}}({\bf r}) =0,                   
\end{equation}
\begin{equation}
\label{2.56}
\mbox{\boldmath $\nabla$}\cdot\hat{\bf D}({\bf r}) = 
\hat{\rho}_{\rm A}({\bf r}),
\end{equation}
\begin{equation}
\label{2.57}
\mbox{\boldmath $\nabla$}\times\hat{\bf H}({\bf r})
- \dot{\hat{\bf D}}({\bf r}) = \hat{\bf j}_{\rm A}({\bf r}),             
\end{equation}
and the operator-valued Newtonian equation of motion 
\begin{equation}
\label{2.58}
\dot{\hat{\bf r}}_\alpha = 
{1\over m_\alpha} \left[ \hat{\bf p}_\alpha 
- q_\alpha \hat{\bf A}(\hat{\bf r}_\alpha) \right],
\end{equation}
\begin{equation}
\label{2.59}
m_\alpha\ddot{\hat{\bf r}}_\alpha = 
q_\alpha \left[\hat{\bf E}(\hat{\bf r}_\alpha) 
+{\textstyle\frac{1}{2}}\left(
\dot{\hat{\bf r}}_\alpha \times \hat{\bf B}(\hat{\bf r}_\alpha) - 
\hat{\bf B}(\hat{\bf r}_\alpha) \times \dot{\hat{\bf r}}_\alpha
\right)\right].
\end{equation}
are fulfilled. In Eq.~(\ref{2.57}),
the atomic current density $\hat{\bf j}_{\rm A}({\bf r})$ reads
\begin{equation}
\label{2.62}
\hat{\bf j}_{\rm A}({\bf r}) = 
{\textstyle\frac{1}{2}} 
\sum_\alpha q_\alpha 
\left[ \delta({\bf r}-\hat{\bf r}_\alpha)\,,\,  
\dot{\hat{\bf r}}_\alpha\right]_+ ,
\end{equation}
where $[\,\,,\,\,]_+$ denotes the anticommutator. Note that
compared with Eqs.~(\ref{2.32}) and (\ref{2.34}), the electric and
displacement fields
now contain additional longitudinal parts that result   
from the charge distribution $\hat{\rho}_{\rm A}({\bf r})$, i.e.,
\begin{equation}
\label{2.60}
\hat{\bf E}({\bf r})  =
\hat{\bf E}_{\rm m}({\bf r}) - 
\mbox{\boldmath $\nabla$}\hat{\varphi}_{\rm A}({\bf r})
= \left[ \int\limits_0^\infty {\rm d}\omega \, 
\underline{\hat{\bf E}}({\bf r},\omega) + {\rm H.c.} \right]
-  \mbox{\boldmath $\nabla$}\hat{\varphi}_{\rm A}({\bf r}) \,,
\end{equation}
\begin{equation}
\label{2.61}
\hat{\bf D}({\bf r}) = 
\hat{\bf D}_{\rm m}({\bf r}) - 
\varepsilon_0\mbox{\boldmath $\nabla$}\hat{\varphi}_{\rm A}({\bf r})
= \left[ \int\limits_0^\infty {\rm d}\omega \,
\underline{\hat{\bf D}}({\bf r},\omega) + {\rm H.c.} \right]
-\varepsilon_0\mbox{\boldmath $\nabla$}\hat{\varphi}_{\rm A}({\bf r}) \,.
\end{equation}
The Maxwell equations (\ref{2.54}) and (\ref{2.56}) simply result 
from the definition of the field operators 
$\hat{{\bf B}}({\bf r})$ [Eqs.~(\ref{2.30}) and (\ref{2.33})] and 
$\hat{{\bf D}}({\bf r})$ [Eqs.~(\ref{2.31}), (\ref{2.52}), and (\ref{2.61})]
respectively. 
The other Maxwell equations (\ref{2.55}) and (\ref{2.57}) and the Newtonian
equation of motion (\ref{2.58}) and (\ref{2.59}) follow from the 
Heisenberg equation of motion, Eq.~({2.37}), with the
Hamiltonian $\hat{H}$ from Eq.~(\ref{2.51}). 


\subsection{The multipolar-coupling Hamiltonian}

In the minimal-coupling scheme, 
the interaction Hamiltonian (\ref{2.63}) is expressed 
in terms of the potentials of the medium-assisted
electromagnetic field.
With regard to (localized) atomic systems (atoms,
molecules etc.) the interaction is commonly desired to be treated
in terms of the electromagnetic field strengths and the
atomic polarization and magnetization. This can be achieved 
by means of a unitary transformation.

Let us consider an atomic system localized at position ${\bf r}_{\rm A}$
and introduce the atomic polarization
\begin{equation}
\label{2.65}
\hat{{\bf P}}_{\rm A}({\bf r}) = 
\sum_\alpha q_\alpha \left(\hat{{\bf r}}_{\alpha}- {\bf r}_{\rm A}\right) 
\int\limits_0^{1} {\rm d}\lambda\, 
\delta\!\left[{\bf r}-{\bf r}_{\rm A}-\lambda\left(\hat{{\bf r}}_\alpha
-{\bf r}_{\rm A}\right)\right], 
\end{equation}
so that the charge density (\ref{2.53}) can be rewritten as
\begin{equation}
\label{2.64}
\hat{\rho}_{\rm A}({\bf r}) = 
\sum_\alpha q_\alpha 
\delta({\bf r}-\hat{\bf r}_{\rm A}) 
- \mbox{\boldmath $\nabla$}\cdot\hat{{\bf P}}_{\rm A}({\bf r}).
\end{equation}
In order to perform the transition from the minimal-coupling scheme 
to the multipolar-coupling scheme, we apply to the variables the unitary 
operator 
\begin{equation}
\label{2.66}
\hat{U} = 
\exp\!\left[\frac{i}{\hbar}\int {\rm d}^3{\bf r}\,
\hat{{\bf P}}_{\rm A}({\bf r})\cdot\hat{\bf A}(\hat{\bf r})\right].
\end{equation}
It is not difficult to prove that the following transformation
rules are valid:
\begin{equation}
\label{2.67}
\hat{\bf r}_\alpha' 
= \hat{U} \hat{\bf r}_\alpha \hat{U}^\dagger = \hat{\bf r}_\alpha,
\end{equation}
\begin{equation}
\label{2.68}
\hat{\bf p}_\alpha' = \hat{U} \hat{\bf p}_\alpha \hat{U}^\dagger 
= \hat{\bf p}_\alpha  - q_\alpha \hat{\bf A}(\hat{\bf r}_\alpha)
- \int {\rm d}^3{\bf r}\,\hat{{\bf n }}_{\alpha}({\bf r}) \times
\hat{{\bf B }}({\bf r}), 
\end{equation}
\begin{eqnarray}
\label{2.69}
\hat{\bf f}'({\bf r},\omega) 
\hspace{-1ex}&=&\hspace{-1ex} 
\hat{U} \hat{\bf f}({\bf r},\omega) \hat{U}^\dagger
\nonumber\\
&=&\hspace{-1ex} 
\hat{\bf f}({\bf r},\omega)
- \frac{i}{\hbar} \sqrt{\frac{\hbar}{\pi\varepsilon_0} 
\varepsilon''({\bf r},\omega)}\, 
\frac{\omega}{c^2}
\int {\rm d}^3{\bf r}'\, \hat{\bf P}^{\perp}_{\rm A}({\bf r}')\cdot
\mbox{\boldmath $G$}^{\ast}({\bf r}',{\bf r},\omega),
\end{eqnarray}
where the abbreviation
\begin{equation}
\label{2.70}
\hat{{\bf n }}_{\alpha}({\bf r}) = 
q_\alpha \left(\hat{{\bf r}}_{\alpha}-\!{\bf r}_{\rm A}\right)
\int\limits_0^{1} {\rm d}\lambda \,\lambda \,
\delta\!\left[{\bf r}-{\bf r}_{\rm A}
-\lambda\left(\hat{{\bf r}}_{\alpha}-\!{\bf r}_{\rm A}\right)\right]
\end{equation} has been used.
Employing equations (\ref{2.67}) -- (\ref{2.69}), 
we can express the Hamiltonian $\hat{H}$ in Eq.~(\ref{2.51}) as the
unitary transform of a new Hamiltonian $\hat{\cal H}$, 
\begin{eqnarray}
\label{2.71}
\hat{H} &=& \hat{U}\,\hat{\cal H}\,\hat{U}^{\dagger} ,
\end{eqnarray}   
where
\begin{eqnarray}
\label{2.72}
\lefteqn{
\hspace*{-4ex}
\hat{\cal H} = 
\int {\rm d}^{3}{\bf r} \int\limits_{0}^{\infty} {\rm d}\omega\,
 \hbar\omega\,\hat{{\bf f}}^{\dagger}({\bf r},\omega)\!\cdot\!
\hat{{\bf f}}({\bf r},\omega)
+ \sum_\alpha \frac{1}{2m_{\alpha}} \left[\hat{{\bf p}}_{\alpha} \!+\!
\int \!{\rm d}^3{\bf r}\,\hat{{\bf n }}_{\alpha}({\bf r}) \times
\hat{{\bf B }}({\bf r})\right]^2
}
\nonumber\\ && \hspace{2ex}
+\,\frac{1}{2}\int {\rm d}^3{\bf r}\,\hat{\rho}_{\rm A}({\bf r}) 
\hat{\varphi}_{\rm A}({\bf r}) 
+\frac{1}{2\varepsilon_0}\int {\rm d}^3{\bf r}\,
\hat{{\bf P}}^{\perp}_{\rm A}({\bf r})\cdot
\hat{{\bf P}}^{\perp}_{\rm A}({\bf r})
\nonumber\\ && \hspace{2ex}
- \int {\rm d}^3{\bf r}\,\hat{{\bf P}}^{\perp}_{\rm A}({\bf r})\cdot
\hat{{\bf E}}_{\rm m} ({\bf r}) +  
\int {\rm d}^3{\bf r}\,\hat{\rho}_{\rm A}({\bf r})\hat{\varphi}({\bf r}) .   
\end{eqnarray}    
In particular when the charged particles form a neutral
atomic system ($\sum_\alpha q_\alpha$ $\!=$ $\!0$), then
Eq.~(\ref{2.72}) takes the form of     
\begin{eqnarray} 
\label{2.73} 
\lefteqn{
\hspace*{-6ex}
\hat{\cal H} = \int {\rm d}^{3}{\bf r} \int\limits_{0}^{\infty}
{\rm d}\omega\, \hbar\omega\,
\hat{{\bf f}}^{\dagger}({\bf r},\omega)\!\cdot\!   
\hat{{\bf f}}({\bf r},\omega) 
+\sum_\alpha \frac{1}{2m_{\alpha}} \left[\hat{{\bf p}}_{\alpha} \!+\!
\int\! {\rm d}^3{\bf r}\,\hat{{\bf n }}_{\alpha}({\bf r}) \times
\hat{{\bf B }}({\bf r})\right]^2
}
\nonumber\\ &&\hspace{2ex}
+\,\frac{1}{2\varepsilon_0}\int {\rm d}^3{\bf r}\,
\hat{{\bf P}}_{\rm A}({\bf r})\cdot\hat{{\bf P}}_{\rm A}({\bf r})
-\int {\rm d}^3{\bf r}\,\hat{{\bf P}}_{\rm A}({\bf r})\cdot
\hat{{\bf E }}_{\rm m}({\bf r}). 
\end{eqnarray}   
Note that the last term on the right-hand side in Eq.~(\ref{2.73})
describes the interaction  between the atomic polarization and the
whole medium-assisted electric field. Introducing the polarization of
the medium 
\begin{equation}
\label{2.73a} 
\hat{{\bf P}}_{\rm m} = \hat{{\bf D}} - \varepsilon_0 \hat{{\bf E}} 
=\hat{\bf D}_{\rm m} - \varepsilon_0 \hat{\bf E}_{\rm m}
= \hat{{\bf D}}_{\rm m}^{\perp} - \varepsilon_0 \hat{{\bf E}}_{\rm m} 
\end{equation}
[see Eqs.~(\ref{2.60}) and (\ref{2.61}] and recalling that 
$\hat{{\bf D}}_{\rm m}^{\perp}$ $\!=$
$\!\hat{{\bf D}}^{\perp}$, we may rewrite the last term in the transformed
Hamiltonian (\ref{2.73}) to obtain
\begin{equation} 
\label{2.74}
-\int \!{\rm d}^3{\bf r}\,\hat{{\bf P}}_{\rm A}({\bf r})\cdot
 \hat{{\bf E }}_{\rm m}({\bf r}) 
= -\frac{1}{\varepsilon_0}
\int \!{\rm d}^3{\bf r}\,\hat{{\bf P}}_{\rm A}({\bf r})\cdot
\hat{{\bf D}}^{\perp}({\bf r})
+\frac{1}{\varepsilon_0} \int \!{\rm d}^3{\bf r}\, 
\hat{{\bf P}}_{\rm A}({\bf r})\cdot\hat{{\bf P}}_{\rm m}({\bf r}).
\end{equation}    

Combining Eqs.~(\ref{2.71}), (\ref{2.73}), and (\ref{2.74}), we
express in the original Hamiltonian the old variables in terms of 
the new ones. The result is the multipolar-coupling Hamiltonian
\begin{eqnarray}
\label{2.75}
\lefteqn{
\hat{H} = \int {\rm d}^{3}{\bf r} \int\limits_{0}^{\infty} {\rm d\omega}
\,\hbar\omega\,\hat{{\bf f}}^{'\dagger}({\bf r},\omega)
\!\cdot\!\hat{{\bf f}}'({\bf r},\omega)
+ \sum_\alpha \frac{1}{2m_{\alpha}} \left[
\hat{{\bf p}}'_{\alpha} \!+\! 
\int\! {\rm d}^3{\bf r}\,\hat{{\bf n}}_{\alpha}({\bf r}) 
 \times \hat{{\bf B }}({\bf r})\right]^2
}
\nonumber\\&& 
+\,\frac{1}{2\varepsilon_0}\int {\rm d}^3{\bf r}\hat{{\bf P}}_{\rm A}({\bf r})
\cdot\hat{{\bf P}}_{\rm A}({\bf r})   
-\,\frac{1}{\varepsilon_0}
\int {\rm d}^3{\bf r}\, \hat{{\bf P}}_{\rm A}({\bf r})\cdot
\hat{{\bf D}}^{'\perp}({\bf r}) 
 +\frac{1}{\varepsilon_0}
\int {\rm d}^3{\bf r} \,\hat{{\bf P}}_{\rm A}({\bf r})
\cdot\hat{{\bf P}}_{\rm m}({\bf r}),
\nonumber\\&&
\end{eqnarray}   
where
\begin{equation}
\label{2.75a}
\hat{\bf D}^{'\perp}({\bf r}) = \hat{\bf D}^\perp({\bf r})
+ \hat{\bf P}_{\rm A}^\perp({\bf r}).
\end{equation}
{F}rom Eq.~(\ref{2.75}) the interaction Hamiltonian is seen to be
\begin{eqnarray}
\label{2.76}
\lefteqn{
\hat{H}_{{\rm int}'} = 
-\frac{1}{\varepsilon_0}
\int {\rm d}^3{\bf r}\,\hat{{\bf P}}_{\rm A}({\bf r})\cdot
\hat{{\bf D}}^{'\perp}({\bf r}) 
 +\frac{1}{\varepsilon_0}
\int {\rm d}^3{\bf r}\,\hat{{\bf P}}_{\rm A}({\bf r})\cdot
\hat{{\bf P}}_{\rm m}({\bf r})
} 
\nonumber\\&&
+\sum_\alpha \frac{1}{2m_{\alpha}} \left[\hat{{\bf p}}'_{\alpha} , 
\int {\rm d}^3{\bf r}\,\hat{{\bf n }}_{\alpha}({\bf r}) \times
\hat{{\bf B }}({\bf r})\right]_+  
+\sum_\alpha \frac{1}{2m_{\alpha}} \left[ 
\int {\rm d}^3{\bf r}\,\hat{{\bf n }}_{\alpha}({\bf r}) \times
\hat{{\bf B }}({\bf r})\right]^2
\nonumber\\&&
\end{eqnarray}    
The first term on the right-hand side in Eq.~(\ref{2.76}) describes
the interaction of the polarization of the atomic system with the
transverse part of the overall displacement field [cf. Eq.~(\ref{2.75a})].
The second term is a contact term  between the medium polarization and the
polarization of the atomic system. The last two terms refer to
magnetic interactions.


\section{Summary and outlook}
\label{summary}

We have developed a general theory of the interaction of the quantized 
electromagnetic field with atoms in the presence of dispersing 
and absorbing dielectric bodies of given Kramers--Kronig consistent 
permittivities. The concept is based on a source-quantity 
representation of the electromagnetic field, in which the 
electromagnetic-field operators are expressed in terms of a continuous
set of fundamental bosonic fields via the Green tensor of the
classical problem. The theory, which is a natural extension of 
the standard concept of mode decomposition, gives a unified
approach to the atom-field interaction, without any restriction
to a particular frequency range.
  
The formalism has been applied quite
recently to the problem of the spontaneous decay of an excited
atom in the presence of dielectric bodies \cite{Ho00}.
In particular, it has been shown that the temporal evolution of
the atomic upper-state-probability amplitude $C_u(t)$ obeys the
integral equation
\begin{equation}
\label{2.77}
C_u(t) = 1+ \int\limits_0^t {\rm d}t'\,\bar K(t-t')\,C_u(t'),
\end{equation}
where the kernel function is determined by the Green tensor
(at the atomic position ${\bf r}_{\rm A}$) as follows:
\begin{equation}
\label{2.78}
K(t-t') 
= \frac{\omega_{\rm A}^2\mu_i\mu_j}{\hbar\pi\varepsilon_0c^2}
\int\limits_0^\infty {\rm d}\omega\,
\frac{{\rm Im}\,G_{ij}({\bf r}_{\rm A},{\bf r}_{\rm A},\omega)}
{i(\omega-\omega_{\rm A})}\left[
e^{-i(\omega-\omega_{\rm A})(t-t')} - 1\right]
\end{equation}
($\omega_{\rm A}$, transition frequency;  
$\mu_i$, transition dipole moment). It is worth noting 
that the integral equation
(\ref{2.77}) applies to the spontaneous decay of an atom
in the presence of an arbitrary configuration of dispersing
and absorbing dielectric bodies. All the matter parameters
that are relevant for the atomic evolution are contained,
via the Green tensor, in the kernel function (\ref{2.78}).
It should be pointed out that the Green tensor has been available
for a large variety of configurations such as
planarly, spherically, and cylindrically
multilayered media \cite{Chew95}. A first evaluation
of the integral equation (\ref{2.77}) for the case of
the atom being placed at the centre of a spherical
micro-cavity whose wall is modeled by a band-gap 
dielectric of Lorentz type has been given in reference 
\cite{Ho00} to which the reader is referred for details.  


\end{document}